\documentclass[12pt,a4paper]{article}
\usepackage[margin=0.9in]{geometry}
\usepackage{graphicx}
\usepackage{hyperref}
\hypersetup{
  colorlinks=true,
  linkcolor=magenta,
  filecolor=magenta,
  urlcolor=cyan,
  citecolor=cyan,
  pdftitle={Electrochemical Phase field Method},
}
\usepackage{graphicx}
\usepackage{dcolumn}
\usepackage[labelfont=bf,labelsep=period]{caption}
\DeclareCaptionFont{mysize}{\fontsize{10}{11}\selectfont}
\usepackage{amsmath,amsfonts,amsthm,amssymb}
\usepackage{bm}
\usepackage{siunitx}
\DeclareSIUnit \amperehour { Ah }
\DeclareSIUnit \molar{\textsc{m}}
\usepackage{sectsty}
\subsectionfont{\itshape}
\providecommand{\p}{\partial}
\providecommand{\dod}[3][]{\dfrac{\mathrm{d}^{#1}#2}{\mathrm{d}{{#3}^{#1}}}}
\providecommand{\dpd}[3][]{\dfrac{\p^{#1}#2}{\p^{{#1}}{{#3}^{#1}}}}
\providecommand{\dfd}[3][]{\dfrac{\delta{^{#1}}#2}{\delta^{{#1}}{{#3}^{#1}}}}

\renewcommand{\v}[1]{\vec{#1}}
\newcommand{\uv}[1]{\hat{#1}}

\renewcommand{\div}[2][]{\v{\nabla}\!{#1}\cdot{#2}}%
\newcommand{\lapl}[1][]{\v{\nabla}^2_{{#1}}}%
\newcommand{\grad}[2][]{\v{\nabla}\!{#1}{#2}}%
\providecommand{\intj}[4][]{\int_{#2}^{#1} \! {#3} \, \mathrm{d} {#4}}
\newcommand\standardstate{{\circ\kern-0.495em-}}
\newcommand{\abs}[2][0]{\lvert {#2} \rvert}
\renewcommand{\text}[1]{\mathrm{#1}}
\graphicspath{{./figures/}}
\begin{document}

\title{Quantitative Phase Field Model for Electrochemical Systems}
\author{Jin Zhang\thanks{Email: jzhang@northwestern.edu}, Alexander F. Chadwick, Peter W. Voorhees}
\date{%
  Department of Materials Science and Engineering, \\Northwestern University, Evanston, IL 60208, USA\\[2ex]
}
\maketitle

\begin{abstract}
  Modeling microstructure evolution in electrochemical systems is vital for understanding the mechanism of various electrochemical processes. In this work, we propose a general phase field framework that is fully variational and thus guarantees that the energy decreases upon evolution in an isothermal system. The bulk and interface free energies are decoupled using a grand potential formulation to enhance numerical efficiency. The variational definition of the overpotential is used, and the reaction kinetics is incorporated into the evolution equation for the phase field to correctly capture capillary effects and eliminate additional model parameter calibrations. A higher-order kinetic correction is derived to accurately reproduce general reaction models such as the Butler-Volmer, Marcus, and Marcus-Hush-Chidsey models. Electrostatic potentials in the electrode and the electrolyte are considered separately as independent variables, providing additional freedom to capture the interfacial potential jump. To handle realistic materials and processing parameters for practical applications, a driving force extension method is used to enhance the grid size by three orders of magnitude. Finally, we comprehensively verify our phase field model using classical electrochemical theory.
\end{abstract}

%% ===========================================================================

\section{Introduction}\label{sec:intro}
Electrochemical systems have many practical applications, such as electrodeposition and electrochemical energy storage. It is believed that electrochemical systems, like batteries, play an essential role in the global effort to achieve carbon neutrality. Understanding detailed electrochemical processes is key for performance optimization, materials and configurations design, and charging and discharging optimization. However, electrochemical systems are heterogeneous and composed of electrodes and electrolytes. Microstructural evolution at the mesoscale, like dendrite growth and isolated-metallic particle formation in batteries, is common in electrochemical processes. Thus, tracking the microstructure evolution is essential for modeling electrochemical processes in electrochemical systems at the mesoscale. 

The phase field method has become an increasingly popular tool for interface tracking due to its flexibility in handling complex morphologies and coupled physical processes \cite{Wang2020}. However, there has been a historical challenge to develop practical and useful phase field models of electrochemical processes, which has led to a wide variety of proposed models. Guyer et al. \cite{Guyer2004,Guyer2004a} developed a variational and thermodynamically consistent model for electrodeposition with linear kinetics that resolves the detailed structure of the electrical double layer. Angstrom scale grid size is needed for this model, so simulations beyond one-dimensional domains are rare. Shibuta et al. \cite{Shibuta2006,Shibuta2007} and Okajima et al. \cite{Okajima2010} developed a variational model based on the classical Kim-Kim-Suzuki model of solidification \cite{Kim1999} but did not explicitly model the electrical double layer. This model allows for a much larger diffuse interface width due to decoupling the bulk and the surface energies. However, a local equilibrium constraint must be fulfilled at each grid point, leading to a high computational cost for general free energy formulations. Cogswell \cite{Cogswell2015} used the grand potential energy to eliminate explicit handling of the local equilibrium constraint. Besides increasing the diffuse interface width, one key advance in the thermodynamically consistent phase field modeling is that the overpotential arises from the variational principle, first given by Bazant \cite{Bazant2013}. Many models employ this definition of the overpotential \cite{Chen2015,Cogswell2015}. Moreover, the general free energy form given by Garc\'ia \textit{et. al.} \cite{Garcia2004} and related variational principles is another important step for phase field models of electrochemical systems.

Another important aspect in quantitative phase field modeling of electrochemical processes is the need to incorporate experimentally accessible reaction kinetics. There are generally two ways to include the reaction kinetics in the governing equations. The first introduces the reaction kinetics as a source term, like in the Allen-Cahn and the Cahn-Hilliard reaction models \cite{Bazant2013,Ely2014,Chadwick2018,Jana2019,Chen2021,Jang2022,Jana2023}. These models enable simulations of a variety of phenomena in electrochemical systems, \textit{e.g.}, as demonstrated by Garc\'ia's group \cite{Jana2019,Jana2023}. However, this approach is nonvariational and thus does not guarantee that the energy will decrease during evolution. Therefore, the resulting model must be carefully verified, especially for the reaction kinetics and the capillary overpotential. Specifically, additional model parameter (phase field mobility) calibration is needed that may not be required in the standard phase field models in order to capture the correct reaction kinetics \cite{Chadwick2018}. A second way is to include the reaction kinetics in the evolution equation for the phase field variable \cite{Cogswell2015,Chen2015}. The source terms come out naturally from the variational derivation of the governing equations in a thermodynamically consistent way. Note that in the work by Chen \textit{et. al.} \cite{Chen2015} and many related works, \textit{e.g.}, \cite{Hong2018,Zhang2021,Ren2022}, the phase field equation is reformulated into an Allen-Cahn reaction equation, so it is not clear if realistic capillary effects are captured and model parameter calibration ($L_\eta$ in \cite{Chen2015}) is still needed. In addition, direct incorporation of the reaction kinetics into the phase field equation may lead to deviation of the phase field kinetics from the desired kinetics at a large overpotential. 

Another aspect that requires careful treatment in developing a phase field model is the potential jump at the electrode-electrolyte interface. The electrostatic potential typically exhibits a large jump (can be more than $1~\si{\volt}$) across the interface due to the existence of the electrical double layer. However, except for models that explicitly resolve the electrical double layer \cite{Guyer2004,Guyer2004a}, the potential jump is frequently overlooked, or a constant jump is assumed, and a single electrostatic potential that continuously varies across the interface is used \cite{Ely2014,Chen2015,Cogswell2015,Hong2018,Jana2019}. 

Finally, to guarantee that the phase field model can provide quantitative predictions, it is important to test whether it can reproduce the existing results of classical electrochemical models. In addition, these tests should be performed with experimentally accessible parameters on realistic length and time scales. However, a comprehensive model verification was not found in the literature.

In this work, we propose a general phase field framework that is fully variational (guarantees energy decrease) and is thermodynamically consistent. We propose a grand potential formulation to decouple the bulk and the surface free energies and eliminate explicitly solving the local equilibrium constraint at each grid point. The classical local equilibrium constraint \cite{Kim1998} is extended using the diffusional electrochemical potential. The variational definition of the overpotential follows from our approach, and the reaction kinetics is considered in the evolution equation for the phase field variable to avoid manual model parameter calibration. A higher-order kinetic correction is derived to reproduce general kinetic models for any overpotential accurately. We consider both electrostatic potentials in the electrode and the electrolyte as independent variables to provide additional freedom to capture the potential jump across the electrode-electrolyte interface. Potential jump varying along the interface can be correctly captured. To handle realistic materials and processing parameters with a micrometer-scale grid size, we apply a driving force extension method \cite{Mapping} to increase the grid size by three orders of magnitude for practical applications. Finally, we provide a comprehensive verification of our model, including the equilibrium behavior, the role of interfacial energy or capillarity, the ability to recover general reaction kinetics, the coupling between diffusion and migration of ions, and time-dependent behavior.

\section{Method}\label{sec:method}

\subsection{Description of the system}\label{sec:method:variables}
\begin{figure}[!t]
  \centering
  \includegraphics[width=.75\textwidth]{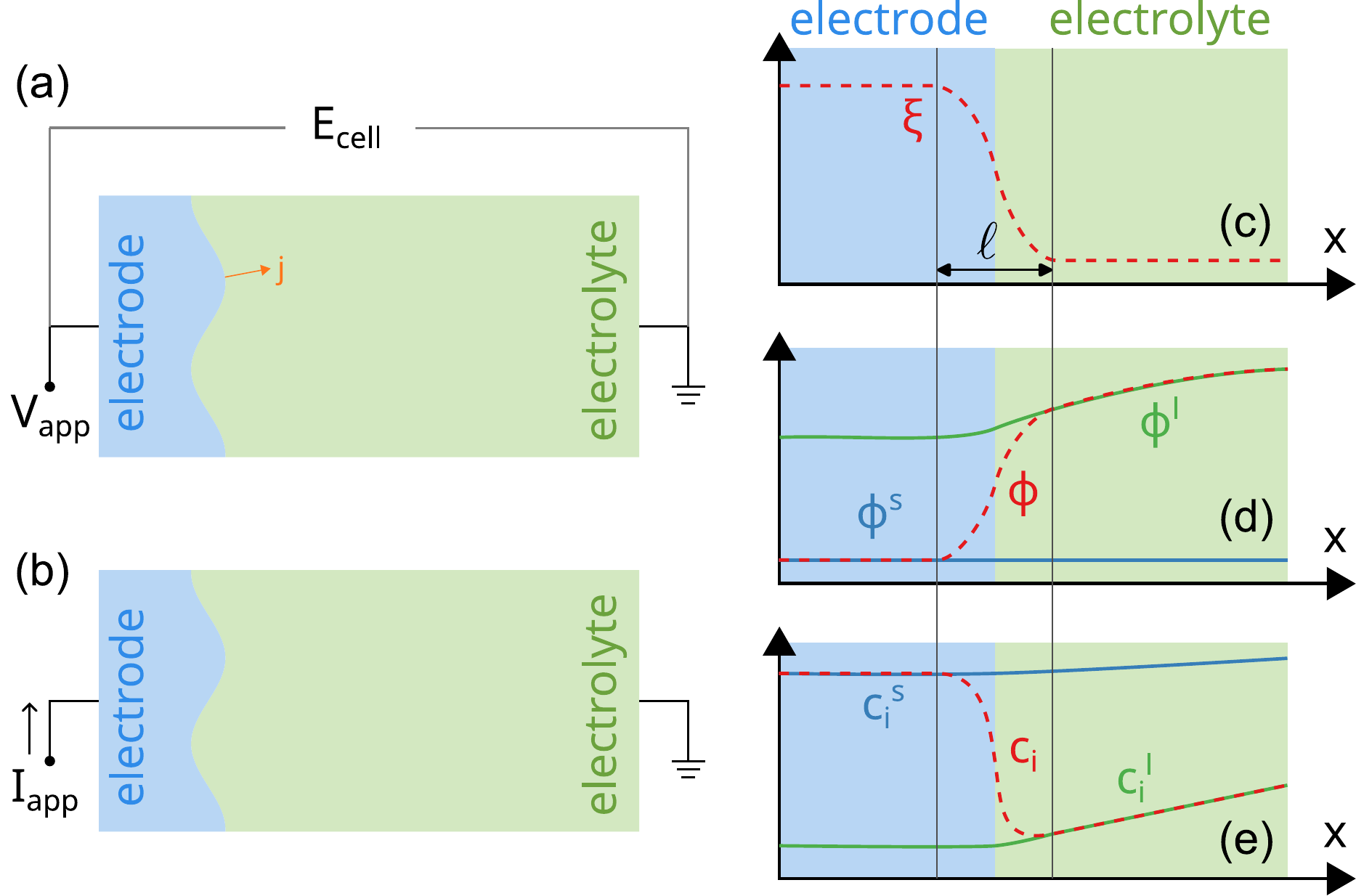}
\caption{Schematic illustration of the electrochemical system for potentiostatic (a) and galvanostatic (b) conditions. $E_{\text{cell}}=-V_{\text{app}}$ is the half-cell voltage. Illustration of field variables to describe the status of the system: (c) the phase field variable $\xi$ varying across a length scale called the diffuse interface width $l$, (d) the electrostatic potential $\phi$ and the virtual electrostatic potentials $\phi^s$ and $\phi^l$ for each phase, and (e) the concentration $c_i$ of species $i$ and the virtual concentrations $c_i^s$ and $c_i^l$ for each phase.}\label{fig:scheme}
\end{figure}
Here we describe variables that characterize the system's status. Consider a system with two phases: a solid phase $s$ (electrode) and a liquid phase $l$ (electrolyte), as shown in Figs.~\ref{fig:scheme}a and b for potentiostatic and galvanostatic loading, respectively. The model is also valid for stress-free, single-phase solid electrolytes without grain boundaries and single ion conduction. Consider the following general reaction at the electrode-electrolyte interface
\begin{equation}\label{eq:reaction}
  \sum_i s_{o,i} O_i + n e^- \leftrightarrow \sum_i s_{r,i} R_i,
\end{equation}
where $O_i$ and $R_i$ are general oxidized and reduced states, respectively, and $s_{o,i}$ and $s_{r,i}$ are corresponding stoichiometric coefficients. The electrode-electrolyte interface will evolve due to the reaction, forming a potentially complex morphology. The location of the interface is described by an implicit function called the phase field variable $\xi$. The phase field variable changes from $1$ in the bulk electrode to $0$ in the bulk electrolyte and has a smooth transition across the interface, as shown in Fig.~\ref{fig:scheme}c. There is an intrinsic length scale $\ell$ associated with the diffuse interface. As shown in Fig.~\ref{fig:scheme}d, the electric potential $\phi$ typically shows a jump at the interface due to the electrical double layer. Here, we focus on systems with a size much larger than the Debye length; therefore, we do not explicitly resolve the double-layer structure. Instead, we define two electrostatic potential fields, $\phi^s$, and $\phi^l$, to implicitly capture the potential jump. As shown in Fig.~\ref{fig:scheme}d, both $\phi^s$ and $\phi^l$ are defined everywhere in the system but have the same value as $\phi$ in the corresponding bulk phases. Following the definition in the Kim-Kim-Suzuki or the grand potential models \cite{Kim1998,Plapp2011}, they are called \textit{virtual} electric potentials, which are only physical in the corresponding phases. Considering an $N$-component system, the concentration of species \textit{i} in phase $\alpha$ is denoted as $c_i^\alpha$. We assume the two phases have the same molar volume $V_m$; in other words, $\sum_{i=1}^N c_i^\alpha=1/V_m, \forall \alpha$. This means only $N-1$ concentrations are independent. By choosing a charge-neutral specie as the $N$-th component (charge number $z_N=0$), we define the diffusional electrochemical potential as $\tilde{\bar{\mu}}_i^\alpha = \bar{\mu}_i^\alpha-\bar{\mu}_N^\alpha$, where $\bar{\mu}_i^\alpha = \mu_i^\alpha + F z_i \phi^\alpha$ is the electrochemical potential, $\mu_i^\alpha$ is the chemical potential, $z_i$ is the charge number and $F$ is the Faraday constant. Concentrations $c_i^s$ and $c_i^l$ are called \textit{virtual} concentrations \cite{Kim1998} (see Fig.~\ref{fig:scheme}e). The total concentration $c_i$, is a combination of $c_i^s$ and $c_i^l$ by a mixture rule $c_i=p(\xi)c_i^s+(1-p(\xi))c_i^l$, where $p(\xi)$ is an interpolation function given below. Generally, there is only one independent concentration \cite{Kim1998,Moelans2011}, so we extend the classical local equilibrium constraint, which requires equal diffusional potential, to the electrochemical system as equal diffusional electrochemical potential $\tilde{\bar{\mu}}_i=\tilde{\bar{\mu}}_i^s=\tilde{\bar{\mu}}_i^l$ to eliminate one extra degree of freedom. However, this local equilibrium constraint must be fulfilled at every grid point in the system, leading to higher computational costs. To solve this problem, we adopt a grand potential formulation \cite{Plapp2011,Cogswell2015} with the diffusional electrochemical potentials $\tilde{\bar{\mu}}_i$ as independent state variables. In this way, the local equilibrium condition is fulfilled naturally. The local equilibrium constraint is used to construct the phase field model and does not mean the interface is at local equilibrium. It should be noted that we use the grand potential formulation to introduce the local equilibrium constraint and simplify the theoretical derivation. However, it is the Helmholtz free energy of the system that is minimized. A total electrostatic potential could be deduced from the equal diffusional electrochemical potential condition and the mixture rule for the total concentration; however, an explicit mixture rule for the total electrostatic potential is unnecessary. In summary, the status of the system can be uniquely described by the phase field $\xi$, the concentrations $\{c_i\}_{i=1}^{N-1}$ (natural variable for the Helmholtz free energy) or the diffusional electrochemical potential $\{\tilde{\bar{\mu}}_i\}_{i=1}^{N-1}$ (natural variable for the grand potential), and the electrostatic potentials $\phi^\alpha,\alpha=s,l$.

\subsection{Basic phase field theory}\label{sec:method:basis:phasefield}
The electrical Helmholtz free energy of the system is
\begin{equation}\label{eq:F}
  \mathcal{F}(\xi,\{c_i\}_{i=1}^{N-1},\phi^s,\phi^l) = \Omega(\xi,\{\tilde{\bar{\mu}}_i\}_{i=1}^{N-1},\phi^s,\phi^l) + \intj{V}{\sum_{i=1}^{N-1} c_i \tilde{\bar{\mu}}_i}{V},
\end{equation}
where the grand potential of the system is defined as
\begin{equation}\label{eq:Omega}
  \Omega(\xi,\{\tilde{\bar{\mu}}_i\}_{i=1}^{N-1},\phi^s,\phi^l) = \intj{V}{\frac{1}{2}\kappa \abs{\grad{\xi}}^2 + m g(\xi) + p(\xi) \omega_e^s + (1-p(\xi)) \omega_e^l}{V},
\end{equation}
where $g(\xi)=\xi^2(1-\xi)^2$ is a double-well potential, $p(\xi)=\xi^2 (3-2\xi)$ is an interpolation function, $\kappa=6 \sigma \ell$ and $m=3\sigma /\ell$ are model parameters related to the surface energy $\sigma$ and the diffuse interface width $\ell$, and \cite{Groot1962,Garcia2004}
\begin{equation}\label{eq:omegae}
  \omega_e^\alpha(\{\tilde{\bar{\mu}}_i\}_{i=1}^{N-1},\phi^\alpha) = \omega^\alpha - \frac{1}{2}\v{D}^\alpha\cdot\v{E}^\alpha, \quad \alpha=s,l,
\end{equation}
where $\omega^\alpha$, $\v{D}^\alpha$ and $\v{E}^\alpha$ are the grand potential density (\si{\joule\per\meter\cubed}), the electrical displacement (\si{\coulomb\per\meter\squared}), and the electric field (\si{\volt\per\meter}) in phase $\alpha$, respectively. Note that in previous models \cite{Shibuta2007,Ely2014,Chen2015,Cogswell2015,Jana2019,Jana2023}, the last term in Eq.~\ref{eq:omegae} is neglected so $\omega_e^\alpha=\omega^\alpha$. Here we keep the general form for consistency. Note here we use a grand potential formulation to decouple the bulk free energy from the surface energy; therefore, it enables the usage of a larger diffuse interface width \cite{Plapp2011,Cogswell2015}.

The second law of thermodynamics requires (see Appendix~A for derivation)
\begin{equation}\label{eq:2ndlaw}
  \dod{\mathcal{F}}{t} = \intj{V}{\dfd{\Omega}{\xi}\dpd{\xi}{t} + \sum_{i=1}^{N-1} \tilde{\bar{\mu}}_i \dpd{c_i}{t} + \sum_{\alpha=s,l}\dfd{\Omega}{\phi^\alpha}\dpd{\phi^\alpha}{t}}{V} \leq 0,
\end{equation}
where the total concentration (\si{\mole\per\meter\cubed}) $c_i = -\delta \Omega/\delta \tilde{\bar{\mu}}_i$ fulfills the mass conservation equation
\begin{equation}\label{eq:conservation:mass}
  \dpd{c_i}{t} = - \div{\v{J}_i},
\end{equation}
where $\v{J}_i$ is the total mass flux (\si{\mole\per\meter\squared\per\second}) of component \textit{i}, which is defined by certain mixture rule from the phase mass flux $\v{J}_i^\alpha$. The charge densities are $\rho^\alpha = \partial \omega^\alpha/\partial \phi^\alpha$. To fulfill the second law of thermodynamics (Eq.~\ref{eq:2ndlaw}), we can have the following constitutive relations from non-equilibrium thermodynamics \cite{Groot1962}
\begin{align}
  \dpd{\xi}{t} &= - \frac{V_m}{6Fl} j(\eta),\label{eq:JXrelation:xi}\\
  \v{J}_i &= - \sum_{j=1}^{N-1} M_{ij} \grad{\tilde{\bar{\mu}}_j},\quad i=1,\cdots,N-1,\label{eq:JXrelation:c}\\
  0 &= \dfd{\Omega}{\phi^\alpha} = \rho^\alpha - \div{\v{D^\alpha}}, \quad \alpha=s,l.\label{eq:JXrelation:phi}
\end{align}
Here we define the variational overpotential \cite{Bazant2013,Cogswell2015}
\begin{equation}\label{eq:overpotential}
  \eta = \frac{V_m}{nF} \dfd{\Omega}{\xi},
\end{equation}
where $n$ is the number of charges transferred (see Eq.~\ref{eq:reaction}). The constant before $j(\eta)$ in Eq.~\ref{eq:JXrelation:xi} merely makes $j$ the interfacial current density (\si{\ampere\per\meter\squared}). Here $j(\eta)$ describes the general reaction in Eq.~\ref{eq:reaction} and can be determined theoretically or experimentally, as long as it fulfills the second law (Eq.~\ref{eq:2ndlaw} and using Eqs.~\ref{eq:JXrelation:xi} and \ref{eq:overpotential}):
\begin{equation}\label{eq:2ndlaw:j}
  - j(\eta) \cdot \eta \leq 0.
\end{equation}
Actually, Eq.~\ref{eq:2ndlaw:j} is fulfilled by typical choices of $j(\eta)$: linear kinetics $j(\eta)=k \eta$ \cite{Guyer2004,Guyer2004a,Shibuta2006,Shibuta2007}, where $k$ is a positive constant, or the Butler-Volmer-like kinetics \cite{Cogswell2015}
\begin{equation}\label{eq:phasefield:j}
  j(\eta) = j_0\left(e^{\alpha_a\frac{F\eta}{RT}} - e^{-\alpha_c\frac{F\eta}{RT}}\right),
\end{equation}
where $j_0$ is the exchange current density, and $\alpha_a$ and $\alpha_c$ are the anodic and cathodic charge transfer coefficients, respectively.

In Eq.~\ref{eq:JXrelation:phi}, we assume the electric field responds to any changes in the system instantaneously. This is valid since the characteristic timescale of the response of the electric field is proportional to the reciprocal of the speed of light, which is much smaller than any characteristic timescale of the interface movement or diffusion. We recover Gauss's law in Eq.~\ref{eq:JXrelation:phi}. For a system size much larger than the Debye length, this equation is typically replaced by the charge neutrality equation $\rho^\alpha=0$ \cite{Dickinson2011} and the charge conservation equation. Actually, the charge conservation equation can be derived by taking the time derivative of Eq.~\ref{eq:JXrelation:phi} and using Amp\'ere's circuital law \cite{Jackson1999} (neglecting magnetic effects):
\begin{equation}\label{eq:conservation:charge}
  \dpd{\rho^\alpha}{t} = \dpd{\div{\v{D}^\alpha}}{t} = \div{\dpd{\v{D}^\alpha}{t}} = - \div{\v{i}^\alpha},
\end{equation}
where $\v{i}^\alpha$ is the current density (\si{\ampere\per\meter\cubed}). Note that this treatment still obeys the second law in Eq.~\ref{eq:2ndlaw}.

The kinetic equation Eq.~\ref{eq:JXrelation:xi}, the diffusion equations Eq.~\ref{eq:conservation:mass} and Eq.~\ref{eq:JXrelation:c} (contain both diffusion and migration of ions), and the charge conservation equations Eq.~\ref{eq:conservation:charge} construct the fundamental equations in our theory. It is worth noting that until this point, no material-specific assumptions, except the energy form given in Eq.~\ref{eq:Omega}, are made. This makes our theory very general and physically rigorous. In the next section, we will consider a particular case of binary electrolyte.

\subsection{Binary electrolyte}\label{sec:method:binaryelectrolyte}
Assume a binary salt dissolves in the solvent $M_{\nu_+}X_{\nu_-}\leftrightarrow \nu_+ M^{z_+} + \nu_- X^{z_-}$. Four species are in the system: cation $+$, anion $-$, solvent $N$, and electron $e^-$. Note we do not need to define the metal as a separate species since it can be viewed as a combination of a cation and an electron. The solvent component $N$ is chosen to define the diffusional electrochemical potential. We assume electronic conduction dominates in the electrode while ionic conduction dominates in the electrolyte. In other words, we treat the electrode as a primarily electric conductor and the electrolyte as a primarily ionic conductor: $c_-^s = c_N^s = c_{e^-}^l=0$. Charge neutrality in each phase gives $z_+ c_+^l + z_- c_-^l = 0$ and $z_+ c_+^s=c_{e^-}^s$. We assume a simple interfacial reaction: $M^{z_+}(l) + n e^{-}(s) \leftrightarrow M (s)$, where $n$ is the number of charges transferred per reaction ($n=z_+$ due to charge conservation). The current density in the electrode equals the electric current density $\v{i}^s = \v{i}_{\text{e}}^s = - \sigma_e^s \grad{\phi^s}$, and the current density in the electrolyte equals the ionic current density $\v{i}^{\,l} = \v{i}_{\text{ion}}^{\,l} = F(z_+\v{J}_+^{\,l}+z_-\v{J}_-^{\,l})$, where $\sigma_e^s$ is the electric conductivity of the electrode. The electrical charge densities in each phase are $\rho_{e}^s = - F c_{e^-}^s = - F z_+ c_+^s$ and $\rho_{e}^l=0$. The ionic charge densities are $\rho_{\text{ion}}^s = F z_+ c_+^s$ and $\rho_{\text{ion}}^l = F (z_+ c_+^l + z_- c_-^l)=0$.

For simplicity, we made the following assumptions: (1) the off-diagonal elements of the mobility matrix are zero $M_{ij}=M_i \delta_{ij}$ (no summation over $i$, $\delta_{ij}$ is the Kronecker delta); (2) the ionic diffusivity in the electrode is much smaller than in the electrolyte $D_+^s\ll D_+^l$; (3) the concentration in the electrode $c_+^s$ is a constant. With these assumptions, the mass conservation equation (Eq.~\ref{eq:conservation:mass} and Eq.~\ref{eq:JXrelation:c}) and the ionic charge conservation equation (Eq.~\ref{eq:conservation:charge}, $\alpha=l$) can be derived as (see Appendix~B for details)
\begin{equation}\label{eq:diffusion}
  (1-p) \dpd{c_+^l}{t} =  \div{\left((1-p)\, D_{\text{eff}}^l \grad{c_+^l}\right)} - (1-p) \frac{\v{i}_{\text{ion}}^{\,l}\cdot\grad{t_+^l}}{z_+F} - p'(\xi)\left((1-t_+^l)c_+^s - c_+^l\right) \dpd{\xi}{t},
\end{equation}
\begin{equation}\label{eq:chargeconservation:ionic}
  -\div{\left( (1-p)\, \v{i}_{\text{ion}}^{\,l} \right)} = p'(\xi) z_+ F c_+^s \dpd{\xi}{t},
\end{equation}
where $t_+^l = z_+^2 M_+^l/(z_+^2 M_+^l + z_-^2 M_-^l)$ is the cation transference number, $D_{\text{eff}}^l = (1-t_+^l) D_+^l + t_+^l D_-^l$ is the (electrolyte) ambipolar/effective diffusivity, and the ionic current in the electrolyte is
\begin{equation}\label{eq:ilion}
  \v{i}_{\text{ion}}^{\, l} = -F z_+\left(D_+^l -D_-^l\right) \grad{c_+^l} - F^2 \left(z_+^2M_+^l + z_-^2M_-^l\right) \grad{\phi^l}.
\end{equation}
The electric charge conservation equation  (Eq.~\ref{eq:conservation:charge}, $\alpha=s$) is
\begin{equation}\label{eq:chargeconservation:electric}
  -\div{\left(p\, \v{i}_{e}^s\right)} = - p'(\xi) z_+ F c_+^s \dpd{\xi}{t}.
\end{equation}

In summary, Eqs.~\ref{eq:JXrelation:xi}, \ref{eq:diffusion}, \ref{eq:chargeconservation:ionic} and \ref{eq:chargeconservation:electric} are the governing equations. Until now, we haven't assumed any form of the Helmholtz free energy density or the relation between $\tilde{\bar{\mu}}_i$ and $c_i^\alpha$. In the next section, We will derive an expression for the overpotential using the free energy for the ideal solution.

\subsection{Ideal solution free energy}\label{sec:method:freeenergy}
To proceed, we need a constitutive relation of the free energy density. For simplicity, we assume an ideal solution of the electrical Helmholtz free energy densities (\si{\joule\per\meter\cubed})
\begin{equation}\label{eq:idealfreeenergy}
  \bar{f}^\alpha(x_+^\alpha, \phi^\alpha) = A^\alpha + B^\alpha x_+^\alpha + \frac{RT}{V_m} \left(x_+^\alpha \ln x_+^\alpha + (1-x_+^\alpha) \ln (1-x_+^\alpha)\right) + \frac{F}{V_m} z_+ x_+^\alpha \phi^\alpha,
\end{equation}
where $x_+^\alpha=V_m c_+^\alpha$ is the molar fraction, $A^\alpha$ and $B^\alpha$ are materials parameters that can be determined experimentally. The grand potential densities $\omega^\alpha$ can be derived from Eq.~\ref{eq:idealfreeenergy} \cite{Plapp2011}. With this, the variational overpotential in Eq.~\ref{eq:overpotential} can be derived as
\begin{equation}\label{eq:dw}
  \eta = \underbrace{\frac{V_m}{nF}\left(-\kappa \lapl{\xi} + m g'(\xi)\right)}_{\text{surface~energy~term}} + p'(\xi) \underbrace{\left(E^\standardstate + \frac{z_+}{n} (\phi^s - \phi^l) + \frac{RT}{nF} \ln \frac{c^s_+}{c^l_+}\right)}_{\text{driving~force~term~} \eta_a},
\end{equation}
where $E^\standardstate = V_m(A^s+B^s-A^l-B^l)/(nF)$ is the standard potential. Here we neglect $1/2\,\v{D}^\alpha\cdot\v{E}^\alpha$ in Eq.~\ref{eq:omegae} since this term is typically orders of magnitude smaller than $\omega^\alpha$. Note that we recover the classical definition of the overpotential \cite{Newman2021}. The derivative of the interpolation function, $p'(\xi)$, can be viewed as a regularized delta function, which merely restricts the quantity to the interface. The surface energy term in Eq.~\ref{eq:dw} is the diffuse interface representation of the capillary overpotential.

\subsection{Relation to classical electrochemical quantities}\label{sec:method:classical}
In the classical electrochemical theory \cite{Bard2001,Newman2021}, the overpotential is only defined at the interface. It can be related to the variational overpotential in Eqs.~\ref{eq:overpotential} and \ref{eq:dw} by integrating along a coordinate $s$ perpendicular to the interface
\begin{equation}\label{eq:classical:overpotential}
  \eta_{\text{classical}} = \frac{1}{6l} \intj[+\infty]{-\infty}{\eta}{s}.
\end{equation}
Similarly, the classical interfacial current density (\si{\ampere\per\meter\squared}) can be related to the corresponding phase field interfacial current density (\si{\ampere\per\meter\squared}):
\begin{equation}\label{eq:classical:current}
  j_{\text{classical}} = \frac{1}{6l} \intj[+\infty]{-\infty}{j(\eta)}{s}.
\end{equation}
The total applied current (\si{\ampere}) can be calculated as
\begin{equation}\label{eq:classical:appliedcurrent}
  I = \frac{1}{6l} \intj{V}{j(\eta)}{V},
\end{equation}
where the integral domain is the whole system. The capacity (\si{\amperehour}) can be calculated as the time integral of $I$: $\intj{}{I}{t}$.

\subsection{Practical aspects}\label{sec:method:other}

\subsubsection{Higher-order kinetic correction}\label{sec:method:correction:kinetic}
For highly nonlinear kinetics or large overpotential, the phase field evolution equation in Eq.~\ref{eq:JXrelation:xi} can deviate from the desired sharp interface result $j_{\text{classical}}=j(\eta_{\text{classical}})$. To improve the accuracy of the phase field kinetics, we expand $j(\eta)$ into higher-order terms and match it with the sharp interface result to introduce the higher-order kinetic correction (see Appendix~C for derivation)
\begin{equation}\label{eq:kinetic:correction}
  -\frac{6Fl}{V_m} \dpd{\xi}{t} = j(\eta) + \sum_{k=2}^{+\infty} p'(\xi) \left(1-p'(\xi)^{k-1}\right) \frac{1}{k!} \left.\dod[k]{j(\eta)}{\eta}\right\vert_{\eta=0} (\eta_a)^k,
\end{equation}
where $\eta_a$ equals the terms in the second bracket in Eq.~\ref{eq:dw}. Note that this kinetic correction is not needed for linear kinetics since all higher-order terms are zero. The higher-order terms are also zero for the particular choice of the interpolation function $p(\xi)=\xi$. However, linear interpolation is generally not preferred as the equilibrium can shift from $\xi=0$ and $\xi=1$ \cite{Wang1993}, the overpotential is not localized near the interface (see Eq.~\ref{eq:dw}), and it is not compatible with the driving force extension given below. In practice, we must truncate the expansion at a finite number $N_k$. The choice of $N_k$ depends on the form of the reaction kinetics and the maximum overpotential. Details will be discussed below. Note that all the even order corrections are zero for Butler-Volmer kinetics with $\alpha_a=\alpha_c=1/2$ and Marcus kinetics. In addition, a stable phase field profile can benefit from this correction at a large overpotential.

\subsubsection{Galvanostatic condition}\label{sec:method:galvanostatic}
The galvanostatic condition (Fig.~\ref{fig:scheme}b) can be achieved by adjusting the applied voltage $\phi^s=V_{\text{app}}$ with a bisection algorithm \cite{DeWitt2015} to match the current calculated from Eq.~\ref{eq:classical:appliedcurrent} to the applied current $I_{\text{app}}$. If the spatial variation of the electric potential of the electrode $\phi^s$ (Eq.~\ref{eq:chargeconservation:electric}) is needed and the conductivity is constant, we can impose a constant charge flux through the Neumann boundary condition to achieve the galvanostatic condition.

\subsubsection{General load profiles}\label{sec:discussion:loading}
The time-dependent loading, either the applied voltage $V_{\text{app}}(t)$ or the applied current $I_{\text{app}}(t)$, can be considered by a temporal discretization $t_k, k=0,1,\cdots$. For each timestep $t_k$ or stage (if a multi-stage time stepper is used), the problem can be treated the same as the case with a fixed applied voltage $V_{\text{app}}(t_k)$ or applied current $I_{\text{app}}(t_k)$. %

\subsubsection{Driving force extension}\label{sec:method:dfex}
The grand potential formulation enables a thick interface width; however, a thick interface is not guaranteed to be numerically stable. The driving force, $\eta_a$, is typically orders of magnitude larger than the surface energy term in Eq.~\ref{eq:dw}. This leads to an unstable phase field profile and restricts the grid size to nanometers or even sub-nanometers. To enable a micrometer grid size, a driving force extension method is used to modify the overpotential in Eq.~\ref{eq:dw} as
\begin{equation}\label{eq:dw:dfe}
  \eta = \frac{V_m}{nF}\left(-\kappa \lapl{\xi} + m g'(\xi)\right) + p'(\xi) \mathcal{P}(\eta_a),
\end{equation}
where $\mathcal{P}$ is the driving force extension operation that maps the driving force $\eta_a$ to a constant perpendicular to the interface. Details on the driving force extension method can be found in \cite{Mapping}. The projection $\mathcal{P}$ keeps the first-order asymptotics of $\eta_a$. Strictly speaking, this numerical treatment is non-variational.

\subsubsection{The limiting current}\label{sec:method:limiting}
The ideal solution free energy given in Eq.~\ref{eq:idealfreeenergy} requires a positive concentration. Numerically solving the equations with an applied charging current density close to the limiting current (typically $>97\%$ of $j_{\text{lim}}$), the virtual concentrations can be negative in the nonphysical region ($c_i^l$ in the electrode). To solve the numerical issue caused by a negative virtual concentration, we introduce a regularization whenever $\ln x_i^\alpha$ is calculated: $\ln x_i^\alpha \rightarrow \ln(\max(\varepsilon,x_i^\alpha))$, where $\varepsilon \ll 1$ is a small positive number. Note that even if the exponential function in the Butler-Volmer equation is combined with the logarithmic function in the overpotential, there are still terms with a negative power of $x_i^\alpha$, forbidding a negative virtual concentration. Another way to handle the limiting current is to correct the Butler-Volmer equation as given in \cite{Chadwick2018}. In this work, we use the simple regularization given above.

It should be noted that charge separation needs to be considered \cite{Bazant2005,Chu2005} to correctly account for the limiting current and over-limiting current behavior. In this case, the charge neutrality assumption employed in this work cannot be used. Gauss's law in Eq.~\ref{eq:JXrelation:phi} should be used instead of the charge neutrality equation to resolve the electrical double layer explicitly. This will be addressed in a future publication.

\section{Results}\label{sec:result}
In this section, we first show the effect of the driving force extension method given above. Then, we provide a comprehensive verification of the proposed general theory, including (a) the equilibrium behavior, (b) the capillary effect, (c) the kinetic reaction, (d) the coupling between diffusion and migration of ions, and (e) the time-dependent behavior. We apply the proposed model to the electrodeposition of lithium metal. The electrolyte is 1M lithium hexafluorophosphate (LiPF$_6$) in ethylene carbonate (EC). We assume the molar fraction of Li$^+$ in the electrode is $x_+^s=0.99$. The temperature is $T=298.15~\si{\kelvin}$. Materials parameters are: charge numbers $z_+=1$, $z_-=-1$, the number of charges transferred $n=1$, the molar volume $V_m = 1.302\times10^{-5}~\si{\meter\cubed\per\mole}$ \cite{Singman1984}, the equilibrium potential of Li$|$Li$^+$ is $\Delta \phi_{\text{eq}} = -1.26~\si{\volt}$ \cite{Liu2021}, from which the free energy related parameters are determined as $A^s=4.5920RT$, $A^l=0$, $B^s=44.4488RT$, $B^l=4.3282RT$ ($R$ is the gas constant), the Li$^+$ diffusivity in electrolyte $D_+^l=3.2\times10^{-10}~\si{\meter\squared\per\second}$ \cite{Valoen2005}, the cation transference number $t_+=0.5$, and the surface energy $\sigma = 1.176~\si{\joule\per\meter\squared}$ \cite{Monroe2003}. For Butler-Volmer kinetics, the exchange current density is $j_0=30~\si{\ampere\per\meter\squared}$ \cite{Monroe2003}, and the symmetric coefficient is $\alpha=0.5$ \cite{Verbrugge1994}. For Marcus and Marcus-Hush-Chidsey kinetics, the exchange current density is $j_0=104~\si{\ampere\per\meter\squared}$, and the reorganization energy is $\lambda=10 RT$ \cite{Boyle2020}. Unless explicitly mentioned, Buter-Volmer kinetics is used. The bulk electrolyte concentration is $c_0=1~\si{\molar}$. The diffuse interface width is $\ell=1.5\Delta x$. In this work, we focus on a half-cell and choose the electric potential on the right side of the electrolyte (see Fig.~\ref{fig:scheme}a) as the reference $\phi^l=0~\si{\volt}$. The electrode is assumed to be equipotential, whose electric potential equals the applied voltage $\phi^s=V_{\text{app}}$. The half-cell voltage is then $E_{\text{cell}} = \phi^l - \phi^s = -V_{\text{app}}$. Numerically, we use the finite difference method for spatial discretization with second-order central differences for the derivatives. Forward-Euler temporal discretization is used for the phase field equation Eq.~\ref{eq:JXrelation:xi}, and backward-Euler is used for the concentration equation Eq.~\ref{eq:diffusion}. The resulting linear systems and the Poisson equations are solved by the Bi-CGSTAB iterative solver \cite{vanderVorst1992} with a multigrid preconditioner.

\subsection{Driving force extension}\label{sec:result:dfe}
\begin{figure}[!t]
  \centering
  \includegraphics[width=.45\textwidth]{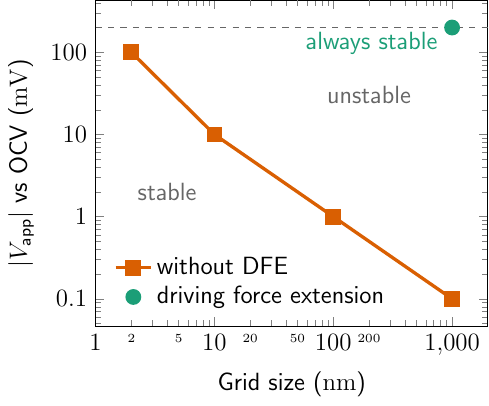}
  \caption{Driving force extension. The solid line shows the maximum applied voltage (versus open-circuit voltage) that gives a stable phase field profile for a given grid size without the driving force extension (DFE). The region above is unstable. With the driving force extension, there is no unstable region. The circular point shows an example.}\label{fig:mapping}
\end{figure}
The surface energy term in Eq.~\ref{eq:dw} can stabilize the phase field profile during evolution. Without the driving force extension in Eq.~\ref{eq:dw:dfe}, the term related to $\eta_a$ typically has a destabilizing effect. To keep a stable traveling wave profile, it is required that these two terms have a similar order of magnitude. The magnitude of the surface energy term is approximately $6 V_m \sigma/(nF \ell)$. Considering a practical overpotential, $\eta_a=200~\si{\mV}$, the maximal interface width is restricted to be $\ell\sim 6 V_m \sigma/(n F \eta_a) =4.8~\si{\nm}$ (this is just a numerical requirement, not a physical one). The grid size will be even smaller to properly resolve the diffuse interface. The maximum applied voltage with a stable phase field profile for a given grid size is shown in Fig.~\ref{fig:mapping}. For practical applied voltages, the stable grid size is on the order of nanometers. This is a significant limitation for simulating practical system sizes that range from micrometers to millimeters. We apply the driving force extension method given above to remove this limitation. As shown in Fig.~\ref{fig:mapping}, for a practical load of $200~\si{\mV}$, the phase field profile is only stable with a $1~\si{\nm}$ grid size without the driving force extension while stable with a $1~\si{\um}$ grid size with the driving force extension. Note that three orders of magnitude increase in grid spacing results in six orders of magnitude increase in timestep size for explicit time stepping. This leads to a significant improvement in computational speed, estimated to be 12 and 15 orders of magnitude for 2D and 3D, respectively, for explicit time stepping.

\subsection{Equilibrium behavior}\label{sec:result:equilibrium}
\begin{figure}[!t]
  \centering
  \includegraphics[width=.45\textwidth]{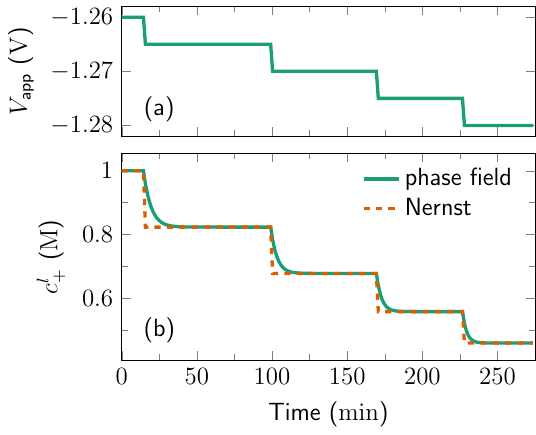}~
  \includegraphics[width=.45\textwidth]{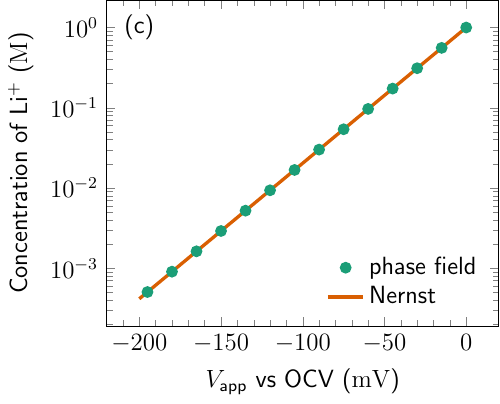}
  \caption{Equilibrium behavior. (a) The applied voltage step history. (b) The Li$^+$ concentration relaxation history compared to the equilibrium concentration determined by the Nernst equation. (c) The equilibrium concentration as a function of the applied voltage (versus open-circuit voltage) compared with the Nernst equation.}\label{fig:verification:nernst}
\end{figure}
We first verify if the proposed model can capture the correct equilibrium behavior. The equilibrium concentration and the cell voltage are related by the Nernst equation \cite{Newman2021}
\begin{equation}\label{eq:verification:nernst}
  E_{\text{cell}} = E_{\text{cell}}^\standardstate - \frac{RT}{F} \ln \frac{c_+^l}{c^\standardstate},
\end{equation}
where $E_{\text{cell}}^\standardstate=-\Delta\phi_{\text{eq}}$ is the equilibrium cell voltage at $c_+^l = c^\standardstate=1~\si{\molar}$. We use step voltage testing on a system given in Fig.~\ref{fig:scheme}a. The system size is $100~\si{\um}$, and the grid size is $1~\si{\um}$. We solve the proposed model in 1D (planar interface) with zero-Neumann boundary conditions for both the phase field and the concentration. For the electrostatic potential, the Dirichlet boundary condition is used on the right $\phi^l=0~\si{\volt}$ and the zero-Neumann boundary condition on the left. Initially, the interface is located in the middle of the system, the electrolyte has a homogeneous concentration $c_+^l=1~\si{\molar}$, and the applied voltage is $V_{\text{app}}=-E_{\text{cell}}^\standardstate$. We then step the applied voltage with a size of $-5~\si{\mV}$, as shown in Fig.~\ref{fig:verification:nernst}a. The interface moves, and the concentration relaxes toward a new equilibrium value determined by the applied voltage; see Fig.~\ref{fig:verification:nernst}b. After complete relaxation (here defined to be the changes in average concentration $c_+^l$ being less than $7.68\times10^{-8}~\si{\molar}$), we record the average $c_+^l$ and the applied voltage $V_{\text{app}}$ and repeat the voltage stepping. The recorded concentrations and applied voltages are plotted in Fig.~\ref{fig:verification:nernst}c and compared with the Nernst equation (solid line). This verifies that our model correctly captures the equilibrium behavior.

\subsection{Capillary effect}\label{sec:result:capillary}
\begin{figure}[!t]
  \centering
  \includegraphics[width=.45\textwidth]{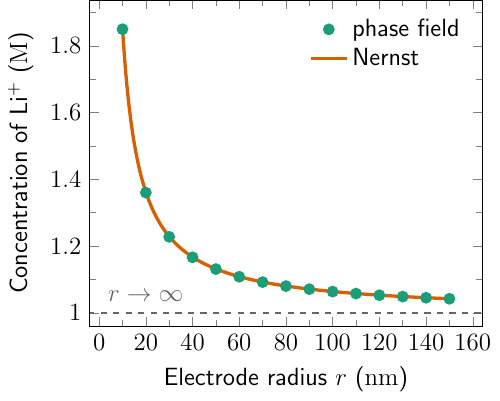}
  \caption{Capillary effect. The equilibrium electrolyte concentration $c_+^l$ as a function of the electrode radius $r$ for an applied potential equals the open-circuit voltage for a planar interface. The gray dashed line shows the equilibrium concentration for a planar interface ($r\rightarrow\infty$).}\label{fig:verification:capillary}%
\end{figure}%
Considering an infinitely long cylindrical electrode with radius $r$, the equilibrium concentration and the cell voltage are related by a modified Nernst equation, accounting for the capillary effect
\begin{equation}\label{eq:verification:nernst:capillary}
  E_{\text{cell}} = E_{\text{cell}}^\standardstate - \frac{RT}{F} \ln \frac{c_+^l}{c^\standardstate} + \frac{V_m}{F} \frac{\sigma}{r}
\end{equation}
To model this problem, we solve the proposed model in 1D cylindrical coordinates. To exaggerate the capillary effect, we model a small cylindrical system of $200~\si{\nm}$ radius with the electrode located at the center and the electrolyte elsewhere. We apply a voltage $V_{\text{app}}=-E_{\text{cell}}^\standardstate$ on the electrode so $E_{\text{cell}} = E_{\text{cell}}^\standardstate$. The equilibrium concentration for a planar interface ($r\rightarrow\infty$) will be $1~\si{\molar}$. Due to the capillary effect, the equilibrium concentration with the cylindrical electrode will deviate from $1~\si{\molar}$. Boundary conditions are the same as in the \textit{Equilibrium behavior} section. The grid size is $\Delta x=0.1~\si{\nm}$. The electrode radius varies from $10~\si{\nm}$ to $150~\si{\nm}$, and the average concentration is measured after full relaxation. A stationary interface is used to fix the electrode radius. This is achieved by freezing the phase field $\xi$. As shown in Fig.~\ref{fig:verification:capillary}, the phase field results compare well with the analytical solution. Moreover, as the electrode radius increases, the equilibrium concentration approaches the equilibrium concentration for a planar interface (shown by the gray dashed line). Note that after full relaxation, the interface will remain stationary even if we let $\xi$ evolve. This verifies that our model captures the correct capillary behavior, critical in many applications such as Li dendrite growth.

\subsection{Kinetic reaction}\label{sec:result:kinetic}
\begin{figure}[!t]
  \centering
  \includegraphics[width=.45\textwidth]{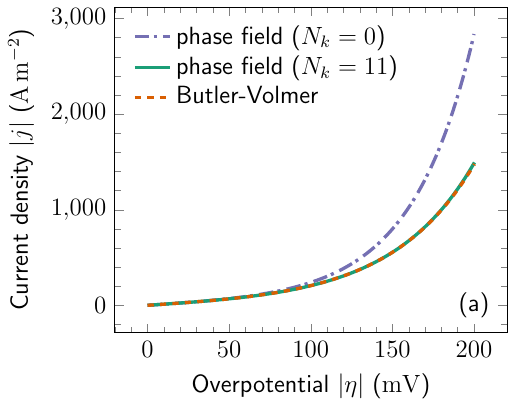}~
  \includegraphics[width=.45\textwidth]{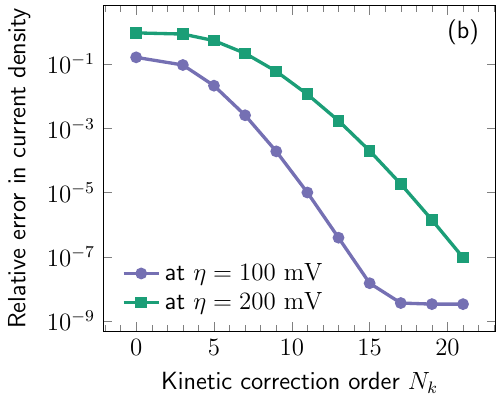}\\
  \includegraphics[width=.45\textwidth]{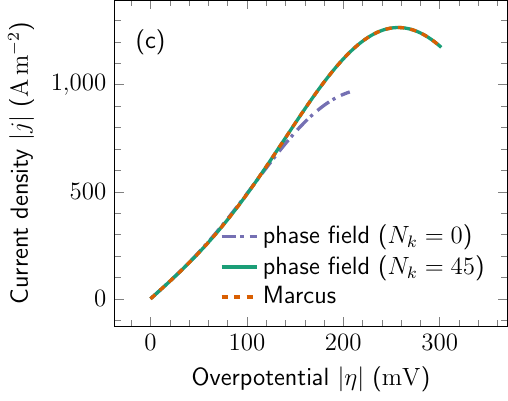}~
  \includegraphics[width=.45\textwidth]{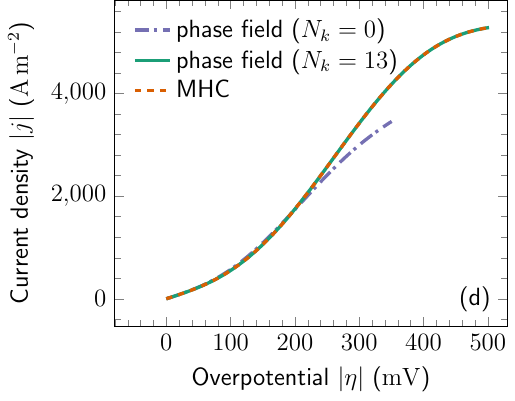}
  \caption{Reaction kinetics. (a) Comparison of the calculated interfacial current density for the Butler-Volmer model without kinetic correction ($N_k=0$) and with kinetic correction truncated at $N_k=11$ with the analytical Butler-Volmer equation. (b) The effect of the kinetic correction order on the relative error in the current density at two overpotential values for the Butler-Volmer case. Comparison of the calculated current density with analytical expression with and without kinetic correction for Marcus (c) and Marcus-Hush-Chidsey (d) models.}\label{fig:verification:kinetics}
\end{figure}
Here, we verify our model with general reaction kinetics using (a) Butler-Volmer equation (Eq.~\ref{eq:phasefield:j}) with $\alpha_a=\alpha_c=1/2$, (b) Marcus kinetics (Eq.~\ref{eq:phasefield:j}) with $\alpha_c=\frac{1}{2} + \frac{F\eta}{4\lambda}$ \cite{Bazant2013} and $\alpha_a=1-\alpha_c$, where $\lambda$ is the reorganization energy \cite{Bard2001}, and (c) Marcus-Hush-Chidsey model \cite{Chidsey1991}. A summary of the equations describing the reaction kinetics for all these models is given in Appendix~D. 

To verify the results for a  broad overpotential range, we manually fix $c_+^l=1~\si{\molar}$ and do not evolve the phase field $\xi$ (can be viewed as in a moving reference frame). We model a planar interface and apply linear potential sweep with a rate of $-1~\si{\mV\per\second}$. The settings are the same as in the \textit{Equilibrium behavior} section. Higher-order kinetic correction given in Eq.~\ref{eq:kinetic:correction} is used but truncated at $N_k$. During the voltage sweeping, the overpotential and current density are calculated from Eqs.~\ref{eq:classical:overpotential} and \ref{eq:classical:current}, respectively, and are shown in Fig.~\ref{fig:verification:kinetics} for both results without the kinetic correction ($N_k=0$) and with a sufficient order of kinetic correction.

For the Butler-Volmer equation, the result without correction (Eq.~\ref{eq:JXrelation:xi}) and with higher-order correction (Eq.~\ref{eq:kinetic:correction}, truncated at $N_k=11$) are compared in Fig.~\ref{fig:verification:kinetics}a. It can be seen that the kinetic correction improves the accuracy for large overpotentials. Figure~\ref{fig:verification:kinetics}b shows the relative error of the current density at two overpotential values as a function of the kinetic correction order $N_k$. The error reduces quickly as more orders of corrections are used. For smaller overpotential, fewer corrections are needed. For the rest of the test cases in this paper, Butler-Volmer kinetics with $N_k=11$ is used. The results for the Marcus and the Marcus-Hush-Chidsey models are shown in Figs.~\ref{fig:verification:kinetics}c and d, respectively. The behavior is similar to the Butler-Volmer case. However, the Marcus kinetics is more nonlinear, so more correction terms are needed. For the Marcus-Hush-Chidsey model, we approximate the Marcus-Hush-Chidsey model by a 13-order polynomial (see Appendix~D), so $N_k=13$ gives the exact result. This verifies that our model can capture general reaction kinetics correctly, even at high overpotentials.

\subsection{Coupling between diffusion and migration}\label{sec:result:coupling}
\begin{figure}[!t]
  \centering
  \includegraphics[width=.45\textwidth]{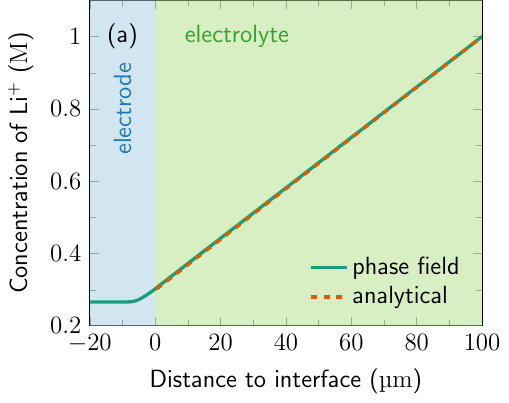}~
  \includegraphics[width=.45\textwidth]{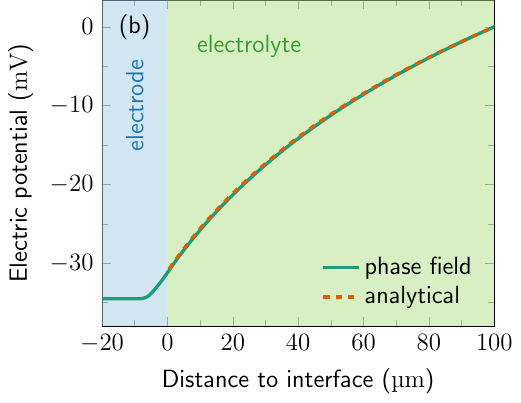}
  \caption{Concentration and potential response of a planar interface to an applied voltage at quasi-steady-state. Comparison of the phase field results with analytical solutions for (a) the concentration field $c_+^l$ and (b) the electrostatic potential field $\phi^l$. Note that the virtual concentration $c_+^l$ has no physical meaning in the electrode, and it is merely a solution to chemical equilibrium.}\label{fig:verification:qss}%
\end{figure}%
To test the coupling between diffusion and migration of ions, we verify our model against the analytical result using a sharp interface model where boundary conditions are applied at the interface,  at quasi-steady-state ($\partial c_+^l/\partial t=0$) for a planar interface. The geometry can be seen in Fig.~\ref{fig:verification:qss}. Dirichlet boundary conditions for the concentration $c_+^l=c_0$ and the electrostatic potential $\phi^l=0$ are used for the right boundary. We apply a voltage $V_{\text{app}} = -E_{\text{cell}}^\standardstate-200~\si{\mV}$ on the electrode. The rest of the settings are the same as the \textit{Equilibrium behavior} section. Governing equations Eqs.~\ref{eq:diffusion}, \ref{eq:chargeconservation:ionic} and \ref{eq:kinetic:correction} are solved in a fully coupled way. To be consistent with the analytical solution, the interface is assumed to be stationary (not evolving $\xi$). The phase field results for the electrolyte concentration $c_+^l$ and the electrostatic potential $\phi^l$ are compared with the analytical solution \cite{Monroe2003} in Figs.~\ref{fig:verification:qss}a and \ref{fig:verification:qss}b, respectively. Note that if a single electrostatic potential field $\phi$ is used, there will be an electrostatic potential jump across the electrode-electrolyte interface from $\phi^l\sim -30~\si{\mV}$ to $\phi^s=-1460~\si{\mV}$, which may be difficult to be resolved accurately in a numerical solver. For higher dimensions, the potential jump will differ from point to point on the interface, which traditional methods overlook. With the two-electrostatic-potential model proposed here, the jump can be correctly handled without explicitly resolving the potential jump. This verifies that the proposed model gives the correct coupling behavior between diffusion and migration of ions in the electrolyte.

\subsection{Transient behavior}\label{sec:result:transient}
\begin{figure}[!t]
  \centering
  \includegraphics[width=.45\textwidth]{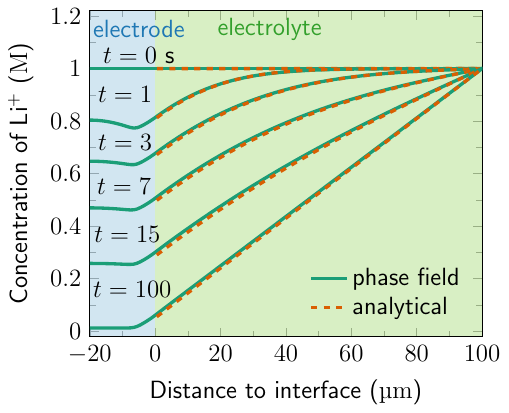}
  \caption{Initial transient of an initially homogeneous electrolyte in response to an applied current of $95\%$ of the limiting current. The phase field calculated electrolyte concentrations $c_+^l$ (solid lines) are compared with the analytical solution (dashed lines) at various times. Note that the virtual concentration $c_+^l$ has no physical meaning in the electrode.}\label{fig:verification:initialtransient}%
\end{figure}%
To examine the time-dependent behavior of the proposed phase field model, we consider the initial transient of a system with a flat electrode-electrolyte interface under galvanostatic conditions. Initially, the system is under open circuit condition, and the electrolyte is homogeneous with a concentration $c_0=1~\si{\molar}$. The geometry is shown in Fig.~\ref{fig:verification:initialtransient}. Boundary conditions are the same as in the previous section. At $t=0$, we apply a constant charging current of $95\%$ of the limiting current $j_{\text{lim}}=2FD_+^lc_0/L=617.5~\si{\ampere\per\meter\squared}$, where $L=100~\si{\um}$ is the system size. With time, the cation depletes close to the interface and eventually reaches a quasi-steady state. The simulated Li$^+$ concentration profile is compared with the analytical solution for the sharp interface model \cite{Monroe2003} at various times in Fig.~\ref{fig:verification:initialtransient}. This indicates that our model can quantitatively predict time-dependent behaviors.

\section{Discussion}\label{sec:discussion}

In summary, the proposed phase field method can successfully model various aspects of electrochemical systems with realistic materials parameters and practical system sizes. The model is fully variational and computationally tractable. Here we discuss possible extensions of the proposed phase field model.

\subsection{Surface energy anisotropy}\label{sec:discussion:anisotropy}
Surface energy anisotropy can be included in the gradient coefficient $\kappa$ in Eq.~\ref{eq:Omega} by letting $\kappa$ depend on the orientation of the interface through the normal to a level-curve of $\xi$, $\uv{n}=-\grad{\xi}/\abs{\grad{\xi}}$. The variational overpotential (Eq.~\ref{eq:overpotential}) then needs to include extra terms related to the derivative of $\kappa$ to $\xi$. This is straightforward and does not change the applicability of the proposed theory and is, therefore, not elaborated here. Details on introducing anisotropy can be found in \textit{e.g.}, \cite{Kobayashi1993,Wheeler1996}. For strong anisotropy, convexification can be employed \cite{Eggleston2001} or the curvature dependence of the interfacial energy can be introduced \cite{Stewart1992,Golovin1998,Siegel2004,Philippe2021}.

\subsection{General free energy forms of the electrolyte solution}\label{sec:discussion:freeenergy}
The proposed theory is formulated to handle general forms of free energy density. The ideal solution given in Eq.~\ref{eq:idealfreeenergy} is used here for simplicity. Consider a general form of the electrochemical potential for concentrated solutions, such as
\begin{equation}\label{eq:concentrated:mu}
  \bar{\mu}_i = \mu_i^\standardstate + RT \ln a_i(\{c_j\}_{j=1}^{N-1}) + z_i F \phi,
\end{equation}
where $a_i$ is the activity that depends on all the independent concentrations $c_i$. The corresponding free energy density can be determined as $f(\{c_i\}_{i=1}^{N-1}) = \sum_{i=1}^{N} c_i \bar{\mu}_i$, from which the grand potential density $\omega$ can be determined. The free energy density can also be defined by the Redlich-Kister expansion \cite{Jana2023}. The derivation of the governing equations and variational overpotential can then proceed similarly.

\subsection{Field-dependent materials parameters}\label{sec:discussion:parameters}
Materials parameters can depend on certain field variables. For example, the ionic diffusivity may depend on the local temperature and concentration. This may introduce additional numerical difficulties, but the proposed theory can be applied to these cases without any change.

\subsection{Other physical phenomena}\label{sec:discussion:physics}
One of the main advantages of the phase field method is that it is easy to add new physics to the model. For example, to account for the stresses in a solid electrolyte, where the strain energy can be included in the total free energy in Eq.~\ref{eq:Omega}, the relevant evolution equations can be derived following the same procedure. If the evolution of the thermal field is needed, the entropy formulation \cite{Wang1993,Bi1998,Sekerka2011} that maximizes the total entropy can be used. To investigate polycrystalline metal anode or the grain structure of solid electrolytes, phase field models with multiple phases \cite{Moelans2011} can be used.

\subsection{SEI effect}\label{sec:discussion:SEI}
SEI can form at the electrode-electrolyte interface, especially for alkali metal anodes. Since the SEI thickness is typically on the nanometer scale, explicitly considering the SEI as a separate phase limits the simulated system size \textit{e.g.} in \cite{Deng2013}. However, the effect of the SEI can be incorporated into the proposed model through parameter modification. For example, the lower ionic diffusivity of the SEI can be included by making the ionic diffusivity dependent on the order parameter \cite{Chen2021}. The stress due to a SEI layer can be incorporated as surface stress. Implicit coupling of the SEI effect into our model will be addressed in a future publication.

\section{Conclusion}\label{sec:conclusion}
In summary, we have introduced a general theory that can be used for quantitative phase field modeling of electrochemical systems that employ realistic materials parameters and practical system sizes. The model is fully variational, guaranteeing that the system's energy must decay during evolution in an isothermal system. The classical definition of the overpotential is recovered for a binary electrolyte and ideal solution free energy. We employ a grand potential formulation to decouple the bulk free energy from the interfacial energy, enabling a much larger diffuse interface width. The classical local equilibrium constraint is extended to the electrochemical case using the diffusional electrochemical potential. In addition, a driving force extension method is applied to increase the grid size by three orders of magnitude and the explicit timestep size by six orders of magnitude compared to standard methods that can be used in a phase field calculation. Thus, the proposed phase field model can be used to efficiently simulate electrochemical systems of practical size with realistic materials parameters.

By incorporating the reaction kinetics in the evolution equation for the phase field parameter, the proposed model can handle general reaction kinetics without additional model parameter calibration. This treatment allows a correct treatment of the capillary overpotential. A higher-order kinetic correction is proposed to accurately handle a range of reaction kinetics models for any overpotential, such as Butler-Volmer kinetics, Marcus kinetics, and Marcus-Hush-Chidsey kinetics.

Both the electrostatic potentials in the electrode and the electrolyte are chosen as independent variables. Compared with previous models where only one electrostatic potential is used, this model allows for more freedom in capturing the correct potential jump at the electrode-electrolyte interface, including a variation of the jump along the interface, without explicitly resolving the potential jump.
 
The model has been verified using several tests: 
\begin{itemize}
\item The model gives the proper equilibrium behavior described by the Nernst equation.
\item The critical role of interfacial energy in the overpotential is captured.
\item General reaction kinetics at the electrode-electrolyte interface are captured.
\item The coupling between diffusion and migration of ions as given by classical theory is recovered.
\item The time evolution of the concentrations in the electrolyte agrees with analytical models.
\end{itemize} 

The proposed phase field framework is very general. It allows applications in a broad range of materials systems, \textit{e.g.}, general electrolyte solution (multicomponent, general charge number), general reaction kinetics, general free energy form for the electrolyte solution (concentrated solutions), and general load profiles (potentiostat, galvanostatic, or time-dependent loading). The nature of the phase field model allows easy extension to include additional physics, like mechanical and thermal effects.

\section*{Acknowledgements}\label{sec:acknowledgement}
This work is sponsored by the Office of Naval Research (ONR) under grant N00014-20-1-2327.

\section*{Appendix}
\appendix
\setcounter{equation}{0}
\renewcommand{\theequation}{A\arabic{equation}}
\section{A. Nonequilibrium thermodynamics}\label{sec:app:net}
The time derivative of Eq.~\ref{eq:Omega} is
\begin{equation}\label{eq:app:dOmega}
  \dod{\Omega}{t} = \intj{V}{\dfd{\Omega}{\xi}\dpd{\xi}{t} + \sum_{i=1}^{N-1}\dfd{\Omega}{\tilde{\bar{\mu}}_i}\dpd{\tilde{\bar{\mu}}_i}{t}+ \sum_{\alpha=s,l}\dfd{\Omega}{\phi^\alpha}\dpd{\phi^\alpha}{t}}{V}.
\end{equation}
The time derivative of Eq.~\ref{eq:F} is (use $c_i=-\delta \Omega/\delta \tilde{\bar{\mu}}_i$)
\begin{align}\label{eq:app:dF}
  \dod{F}{t}&= \dod{\Omega}{t} + \intj{V}{\sum_{i=1}^{N-1} \dpd{c_i}{t} \tilde{\bar{\mu}}_i + c_i \dpd{\tilde{\bar{\mu}}_i}{t}}{V}, \\
  &=\intj{V}{\dfd{\Omega}{\xi}\dpd{\xi}{t} + \sum_{i=1}^{N-1} \tilde{\bar{\mu}}_i \dpd{c_i}{t} + \sum_{\alpha=s,l}\dfd{\Omega}{\phi^\alpha}\dpd{\phi^\alpha}{t}}{V}.
\end{align}

\section{B. Derivation of the evolution equations}\label{sec:app:derivation}
We assume the electrode is equipotential ($\sigma_e^s=\infty$), so the solution of Eq.~\ref{eq:conservation:charge} (for $\phi^s$) is $\phi^s=V_{\text{app}}$. Following the grand potential formulation, the time derivative of the total concentration can be written as
\begin{align}\label{eq:app:dc}
  \dpd{c_+(\xi,\tilde{\bar{\mu}}_+,\phi^s,\phi^l)}{t}&= \dpd{c_+}{\xi} \dpd{\xi}{t} + \dpd{c_+}{\tilde{\bar{\mu}}_+}\dpd{\tilde{\bar{\mu}}_+}{t} + \dpd{c_+}{\phi^s} \dpd{\phi^l}{t} + \dpd{c_+}{\phi^l} \dpd{\phi^l}{t} \nonumber\\
    &=p'(\xi) (c_+^s-c_+^l) \dpd{\xi}{t} + \chi_+ \dpd{\tilde{\bar{\mu}}_+}{t} - z_+ F \left(p\chi_+^s \dpd{\phi^s}{t}+(1-p)\chi_+^l\dpd{\phi^l}{t}\right),
\end{align}
where
\[\chi_+^s = \dpd{c_+^s}{\tilde{\bar{\mu}}_+}, \quad \chi_+^l = \dpd{c_+^l}{\tilde{\bar{\mu}}_+}, \quad \chi_+ = p \chi_+^s + (1-p)\chi_+^l, \quad \dpd{c_+^s}{\phi^s} = -z_+ F \chi_+^s, \quad \dpd{c_+^l}{\phi^l} = -z_+ F \chi_+^l.\]
Substituting into Eq.~\ref{eq:conservation:mass} and using Eq.~\ref{eq:JXrelation:c}, we have
\begin{equation}\label{eq:dmu}
p \chi_+^s \dpd{\tilde{\mu}_+^s}{t} + (1-p) \chi_+^l \dpd{\tilde{\mu}_+^l}{t} = \div{\left(M_+ \grad{\tilde{\bar{\mu}}_+}\right)} - p'(c_+^s-c_+^l) \dpd{\xi}{t}.
\end{equation}
Assuming $D_+^s\ll D_+^l$ and $c_+^s = \text{constant}$, we can simplify Eq.~\ref{eq:dmu} and Eq.~\ref{eq:conservation:charge} (for $\phi^l$) to
\begin{equation}\label{eq:dc:final}
  (1-p) \dpd{c_+^l}{t} = \div{\left((1-p) M_+^l \grad{(\tilde{\mu}_+^l+z_+F \phi^l)}\right)} - p'(c_+^s-c_+^l) \dpd{\xi}{t},
\end{equation}
\begin{equation}\label{eq:phil:final}
p' z_+ c_+^s \dpd{\xi}{t} = \div{\left( (1-p)\left(z_+ (D_+^l - D_-^l) \grad{c_+^l}  + F \left(z_+^2M_+^l + z_-^2M_-^l\right) \grad{\phi^l} \right) \right)},
\end{equation}
where we use $M_+ = p(\xi) M_+^s + (1-p(\xi)) M_+^l = (1-p(\xi)) D_+^l \chi_+^l$. Using Eq.~\ref{eq:phil:final} to eliminate $\phi^l$ in Eq.~\ref{eq:dc:final}, we end up with Eq.~\ref{eq:diffusion}. Note that an anti-trapping current is not needed due to a small interfacial velocity.

\section{C. Derivation of the higher-order kinetic correction}\label{sec:app:kcorr}
From Eq.~\ref{eq:dw}, the classical sharp interface overpotential $\eta_s$ is related to the variational phase field overpotential $\eta$ as $\eta = p' \eta_s$, where $p'$ is the derivative of the interpolation function $p(\xi)$. Taylor expansion of $j(\eta)$ around $\eta=0$ gives
\begin{equation}\label{eq:app:kcorr:j}
  j(\eta) = j(p' \eta_s) = j(0) + \sum_{k=1}^\infty \frac{1}{k!} \left.\dod[k]{j(\eta)}{\eta}\right\vert_{\eta=0} (p' \eta_s)^k.
\end{equation}
The desired kinetic behavior for the phase field model is $p'j(\eta_s)$, which can be Taylor expanded around $\eta_s=0$ as
\begin{align}\label{eq:app:kcorr:js}
  p' j(\eta_s) &= p' j(0) + p' \sum_{k=1}^\infty \frac{1}{k!} \left.\dod[k]{j(\eta)}{\eta}\right\vert_{\eta=0} (\eta_s)^k \nonumber\\
  &= j(\eta) + \sum_{k=1}^\infty \frac{1}{k!} \left.\dod[k]{j(\eta)}{\eta}\right\vert_{\eta=0} \eta^k \left(\frac{p'}{(p')^k}-1\right),
\end{align}
where Eq.~\ref{eq:app:kcorr:j} is used. Eq.~\ref{eq:app:kcorr:js} can be rearranged to give the higher-order kinetic correction in Eq.~\ref{eq:kinetic:correction}.

\section{D. Reaction kinetics}\label{sec:app:reaction}
For simplicity, the overpotential $\eta$ and the reorganization energy $\lambda$ in this section are nondimensionalized by $RT/F$ and $RT$, respectively. The Butler-Volmer equation is \cite{Bard2001}
\begin{equation}\label{eq:app:reaction:BV}
  j(\eta) = j_0 \left(\exp\left(\frac{\eta}{2}\right) - \exp\left(-\frac{\eta}{2}\right) \right).
\end{equation}
The Marcus kinetic equation is \cite{Marcus1965,Bard2001,Bazant2013}
\begin{equation}\label{eq:app:reaction:Marcus}
j(\eta) = j_0 \left(\exp\left(\left(\frac{1}{2}-\frac{\eta}{4\lambda}\right)\eta\right) - \exp\left(-\left(\frac{1}{2}+\frac{\eta}{4\lambda}\right)\eta\right) \right).
\end{equation}
The Marcus-Hush-Chidsey model is \cite{Chidsey1991}
\begin{equation}\label{eq:app:reaction:MHC}
  j(\eta) = \frac{j_0}{A} \left(\int^{+\infty}_{-\infty}\!{\exp\left(-\frac{(x-\lambda + \eta)^2}{4\lambda}\right) \frac{\mathrm{d}x}{1+e^{x}}} - \int^{+\infty}_{-\infty}{\exp\left(-\frac{(x-\lambda - \eta)^2}{4\lambda}\right) \frac{\mathrm{d}x}{1+e^{x}}}\right),
\end{equation}
where
\[A = \intj[+\infty]{-\infty}{\exp\left(-\frac{(x-\lambda)^2}{4\lambda}\right) \frac{1}{1+e^{x}}}{x}.\]
In the practical implementation of the Marcus-Hush-Chidsey model, it is time-consuming to do the numerical integration for each grid point and each timestep. Instead, we use a polynomial fit of the numerical integration result for $\lambda = 10$ within the range of $\abs{\eta}\leq 35$:
\begin{align}\label{eq:app:reaction:MHC:polynomial}
j(\eta) = j_0 &\left(
   20.1354 \left(\frac{\eta}{20}\right)
  +201.893 \left(\frac{\eta}{20}\right)^3
  -370.66 \left(\frac{\eta}{20}\right)^5
  +304.759 \left(\frac{\eta}{20}\right)^7\right.\nonumber\\ 
  &\left.-131.398 \left(\frac{\eta}{20}\right)^9
  +28.8158 \left(\frac{\eta}{20}\right)^{11}
  -2.53411 \left(\frac{\eta}{20}\right)^{13}
\right).
\end{align}
The maximum relative error of the polynomial fit is $1.6\%$. Analytical approximation of the Marcus-Hush-Chidsey kinetics \cite{Zeng2014} can also be used, providing proper calculation of the derivatives.

% References

\end{document}